\documentclass[aps,pra,reprint,groupedaddress,nofootinbib,longbibliography,superscriptaddress]{revtex4-2}
\usepackage{placeins}
\usepackage{amsmath,amsfonts,amssymb}
\usepackage{amsmath}
\usepackage{graphicx}
\usepackage[colorlinks=true, allcolors=blue]{hyperref}
\usepackage{siunitx}
\usepackage{changes}
\usepackage{soul}

\begin{document}

\title{Position measurement of a levitated nanoparticle\\ via 
	interference with its mirror image}

\author{Lorenzo Dania}
\email[]{lorenzo.dania@uibk.ac.at}
\address{Institut f{\"u}r Experimentalphysik, Universit{\"a}t Innsbruck, Technikerstra\ss e 25, 6020 Innsbruck,
	Austria}
\author{Katharina Heidegger}
\address{Institut f{\"u}r Experimentalphysik, Universit{\"a}t Innsbruck, Technikerstra\ss e 25, 6020 Innsbruck,
	Austria}
\author{Dmitry S. Bykov}
\address{Institut f{\"u}r Experimentalphysik, Universit{\"a}t Innsbruck, Technikerstra\ss e 25, 6020 Innsbruck,
	Austria}
\author{Giovanni Cerchiari}
\address{Institut f{\"u}r Experimentalphysik, Universit{\"a}t Innsbruck, Technikerstra\ss e 25, 6020 Innsbruck,
	Austria}
\author{Gabriel Araneda}
\address{Department of Physics, University of Oxford, Clarendon Laboratory, Parks Road, Oxford OX1 3PU, U.K.}
\author{Tracy E. Northup}
\address{Institut f{\"u}r Experimentalphysik, Universit{\"a}t Innsbruck, Technikerstra\ss e 25, 6020 Innsbruck,
	Austria}

\date{\today}

\begin{abstract}
Interferometric methods for detecting the motion of a levitated nanoparticle provide a route to the quantum ground state, but such methods are currently limited by mode mismatch between 
the reference beam and the dipolar field scattered by the particle.
Here we demonstrate a self-interference method to detect the particle's motion that
solves this problem.
A Paul trap confines a charged dielectric nanoparticle in high vacuum, and a mirror retro-reflects the scattered light. We measure the particle's motion with a sensitivity of $\SI{1.7e-12}{\meter/\sqrt{\hertz}}$, corresponding to a detection efficiency of 2.1\%, with a numerical aperture of 0.18. 
As an application of this method, we cool the particle, via feedback, to temperatures below those achieved in the same setup using a standard position measurement.
\end{abstract}

\maketitle
In state-of-the-art levitated optomechanics experiments, dipole radiation elastically scattered by a submicron particle is collected by a lens with high numerical aperture (NA)~\cite{magrini2021,tebbenjohanns2021}, and interference 
with a Gaussian reference beam enables reconstruction of the particle's motion~\cite{tebbenjohanns2019,gittes98}. The detection efficiency achievable with these techniques has two fundamental limits~\cite{tebbenjohanns2019}: first, the particle radiation away from the lens aperture is not collected, and thus information about the particle's position is lost. Second, the dipole radiation wavefront and polarization do not match the Gaussian beam profile, reducing the interferometric visibility and thus the position measurement sensitivity. 
Both the limited collection efficiency and the mode mismatch present a challenge for achieving quantum control of a levitated particle's motion.  While optically levitated particles have been cooled to their motional ground state~\cite{delic2020,magrini2021,tebbenjohanns2021}, the quantum regime has remained out of reach for particles confined in ion traps~\cite{dania2020,millen2015} and magnetic traps~\cite{gieseler2020,tim2019,bradley}, in which optical access is limited by the trap geometry.
These latter approaches are nevertheless of particular interest for quantum-mechanical experiments with macroscopic objects, as they offer an environment free of light-induced decoherence.

Remarkable collection efficiencies for both nanoparticle emitters and atomic ion fluorescence have been achieved with deep parabolic mirrors~\cite{Salakhutdinov16,maiwald12,chou_note}, but such setups are optimized for enhanced light-matter interaction rather than interferometric position detection. Near-ideal mode matching is obtained in experiments in which the dipolar emission is referenced to its mirror image via homodyne detection.
For example, self-interference detection is exploited in trapped-ion experiments for detecting mechanical motion with sensitivities at the level of single quanta~\cite{cerchiari2021b,cerchiari2021c}, and similar techniques are used in fluorescence microscopy~\cite{swan2003,davis07} for determining the structure and diffusion of chemical components in biological tissue with sub-wavelength resolution~\cite{huang2009}. Recently we argued that, compared to known methods, self-homodyne detection is the closest realization of a quantum-limited measurement for the position of a dipolar scatterer~\cite{cerchiari2021}, but until now, this technique has not been adapted to a levitated optomechanics experiment, nor has its efficiency been compared experimentally to that obtained with other methods.

Here we report on the optical detection of the motion of a silica nanoparticle by self-interference, and we compare the detection efficiency with both a forward-detection technique implemented in parallel and ideal quantum-limited detection. Furthermore, we show that the position signal obtained with the self-homodyne method enables real-time feedback cooling of the particle's motion. In our proof-of-principle demonstration, the nanoparticle is trapped in a linear Paul trap under high vacuum, but the detection technique can be extended to setups that confine dipolar emitters by magnetic or optical forces.    
\begin{figure}[ht]
	\centering
	\includegraphics[width=1\linewidth]{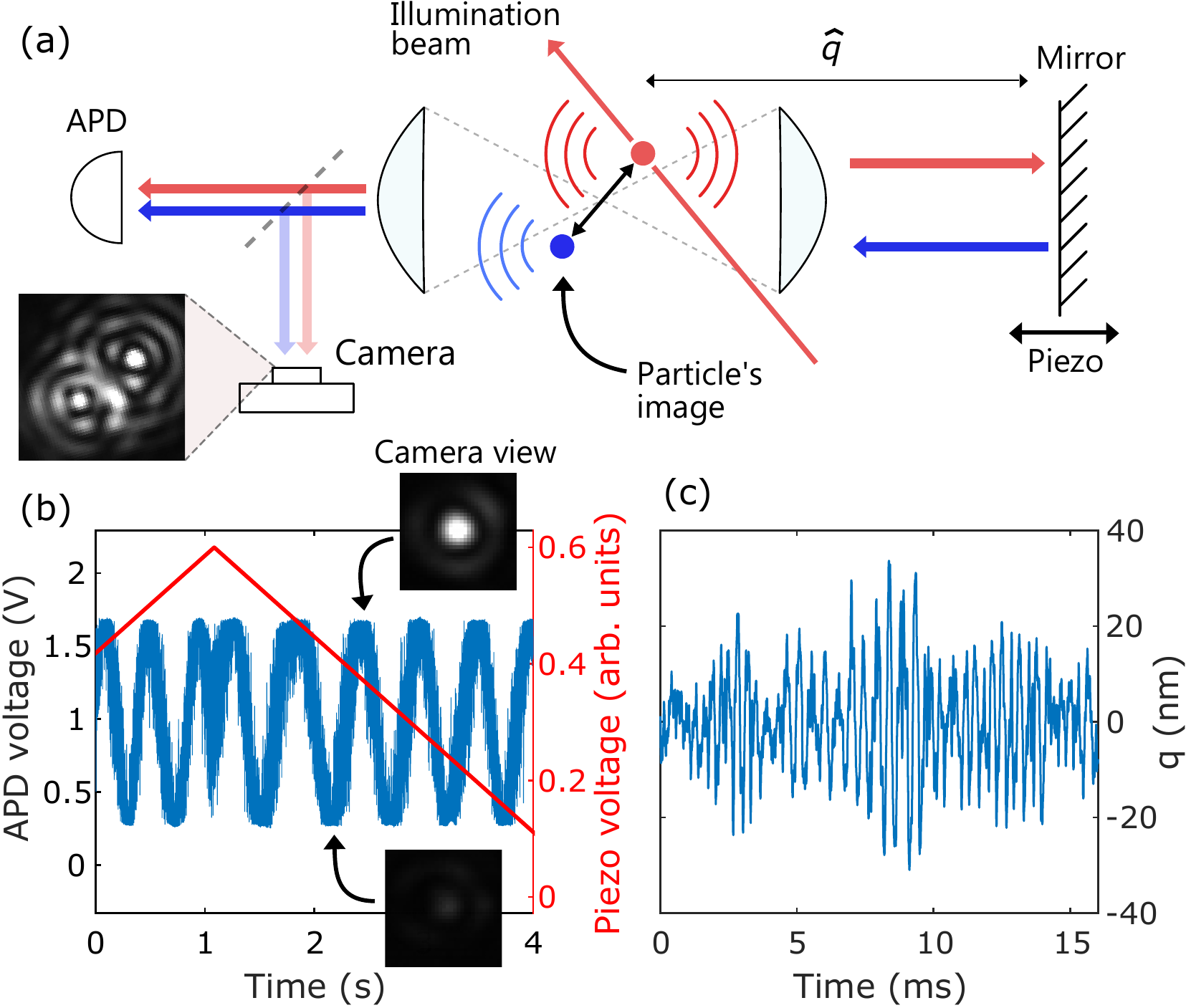}
	\caption{Basic principles of the self-homodyne technique. (a) The particle is illuminated by a laser beam and two confocal lenses collect the scattered light. The image created by one lens is reflected back to the particle by a distant flat mirror. The reflected light interferes with the directly scattered light, and the other lens images the two fields on a detector. The phase of the interferometric signal measured by the detector depends on the mutual particle--mirror distance along the $\hat{q}$ axis. Inset shows a camera image of the nanoparticle and its mirror image. Here, for visualization purposes, the images are displaced from each other. (b) Typical detector output (blue) when the mirror is linearly displaced in time with a piezo translation stage. The driving ramp voltage of the mirror (red) occurs with a sub-hertz period. Insets show camera images obtained at maximum and minimum of the interferometric fringe, respectively. (c) Time trace of the particle's displacement $q$ for the case of a fixed mirror position. Particle oscillations occur at frequencies of a few kilohertz.}
	\label{fig:fig_1}
\end{figure}
\begin{figure}[ht]
	\centering
	\includegraphics[width=1\linewidth]{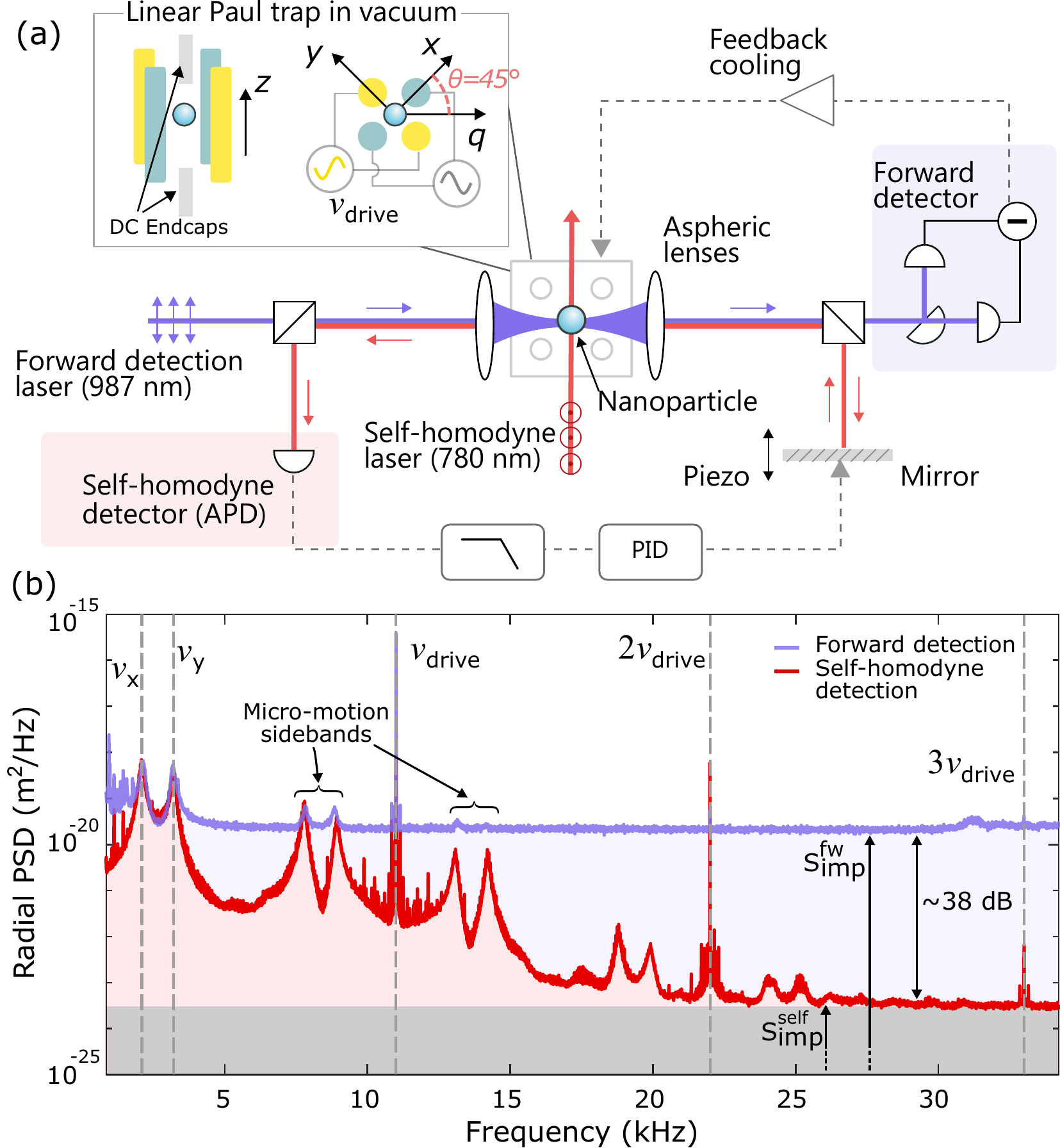}
	\caption{Simultaneous self-homodyne and forward detection of a levitated nanoparticle. (a) Schematic overview of the setup. Lasers for the two detection methods are decoupled using polarization optics. For self-homodyne detection, the particle is illuminated with a weakly focused beam, and the particle's motion along the detection axis $\hat{q}$ is detected by an avalanche photodiode (APD). The APD signal is filtered and sent to feedback electronics that stabilize the mirror position. For forward detection, light propagating along $\hat{q}$ is focused onto the particle and then guided, together with the forward-scattered light from the particle, to a set of balanced photodiodes. The light scattered by the particle from the two lasers is collected by a pair of aspheric lenses (NA$=0.18$), which also focus and collimate the $\SI{987}{\nano\meter}$ beam. The inset shows the Paul trap axes. (b) Power spectral density (PSD) of the radial motion of one particle acquired simultaneously with the forward detection method (red) and with the self-homodyne method (blue). Dashed gray lines indicates the trap radial frequencies $\nu_x$ and $\nu_y$, the Paul trap's drive frequency $\nu_{\mathrm{drive}}$, and the first two harmonics of $\nu_{\mathrm{drive}}$  The measurement imprecisions $S_{\mathrm{imp}}^{\mathrm{self}}$ and $S_{\mathrm{imp}}^{\mathrm{fw}}$ are indicated.
	}
	\label{fig:fig_2}
\end{figure}
The self-homodyne method allows us to measure the position of a nanosphere relative to a mirror.  The essence of self-homodyne can be understood by considering a movable mirror, a detector, and a levitated particle illuminated by a laser beam (Fig.\ref{fig:fig_1}a).  The particle is located between the mirror and the detector.  Laser light scattered by the particle may impinge directly on the detector, or it may be reflected first from the mirror and only afterwards reach the detector.  Here, we consider a planar mirror and a planar detector, so we also include two confocal lenses that collimate the scattered light.  These two light fields --- the one that is directly scattered, and the one that is first reflected --- interfere at the detector; the interference phase depends on the distance between particle and mirror (Fig.\ref{fig:fig_1}b). In contrast, if the mirror position is held fixed, one observes a signal proportional to the particle position (Fig.\ref{fig:fig_1}c).

We employ two different modes to control the mirror's position via a piezoelectric stage: either we actively stabilize the mirror's position in order to monitor the particle, or we apply a linear ramp to calibrate the amplitude of the particle's motion.  In the first case, the mirror position is locked to a point at which the fringes in Fig.\ref{fig:fig_1}b have a maximum slope, so that we have both linear detection and the best sensitivity.  The lock compensates for drifts in the mirror position; it is able to distinguish between changes in the mirror position and changes in the nanosphere's position because the former occur on time scales much longer than the nanosphere period of oscillation.  Note that the lock will only work if the nanosphere's amplitude of motion $\Delta q$ is below $\lambda/4$, that is, within the turning points of the error signal, and linear detection requires the stronger condition $\Delta q \ll \lambda/4$.  
	
In the second case, when we are interested in calibrating the nanosphere's amplitude of motion along the axis $\hat{q}$ of Fig.\ref{fig:fig_1}a, we ramp the voltage applied to the piezo and measure the signal $V(t)$ recorded by the detector.  In the linear regime around the setpoint, for low values of the NA, we can convert $V(t)$ to a displacement $q(t)$ along $\hat{q}$ using the expression $q(t) = V(t)/S$, where $S = 4\pi A/\lambda$ is the maximum slope of the interference fringe and $A$ is the fringe amplitude~\cite{eschner2001}. 
A low NA ensures that any phase change induced by the ramping mirror cannot be distinguished from a displacement of the particle along $\hat{q}$. The more general case for any NA is derived and discussed in the Supplemental Material. 
This calibration relies only on the wavelength of the illuminating field; in contrast to other calibration techniques \cite{Hebestreit2018}, no prior knowledge is required of the particle's mass or charge, and the particle does not need to be in thermal equilibrium with the background gas.
For the measurements described below, the calibration is performed under high vacuum.

Having examined the core components, we now consider the larger experimental apparatus (Fig.~\ref{fig:fig_2}a).
The nanoparticle, a silica sphere $\SI{300}{\nano\meter}$ in diameter\footnote{Bangs Laboratories, Inc.}, is confined in a linear Paul trap \cite{bykov2019} under pressures as low as $\SI{2e-8}{\milli\bar}$.
The trap drive frequency is $\nu_{\mathrm{drive}}=\SI{11}{\kilo\hertz}$,
and the secular trap frequencies are $(\nu_x, \nu_y, \nu_z) = (2.1, 3.2, 1.1)\,\si{\kilo\hertz}$, where $\hat{z}$ is the trap axis.
Self-homodyne detection is implemented with a $\SI{780}{\nano\meter}$ laser beam. The beam is polarized along the $\hat{z}$ axis and illuminates the nanoparticle in the direction perpendicular to the mirror-particle axis $\hat{q}$; the beam waist at the trapping position is $\SI{0.29(1)}{\milli\meter}$. Due to the Paul trap geometry, the self-homodyne detection axis $\hat{q}$ intersects both the $\hat{x}$ and the $\hat{y}$ axes at $\theta=\SI{45}{\degree}$, enabling displacement detection along both axes with equal sensitivity.
When a nanoparticle is loaded into the Paul trap, it has an amplitude of motion larger than an optical wavelength.
To cool the particle into the regime in which self-homodyne detection operates, we implement an auxiliary detection scheme.
A $\SI{987}{\nano\meter}$ laser beam is focused on the particle with a waist of \SI{2.7(3)}{\micro\meter} along the $\hat{q}$ axis; its polarization is perpendicular to the $\hat{z}$ axis. 
We detect forward-scattered light from this beam in a split detection scheme, which we refer to as forward detection \cite{tebbenjohanns2019,dania2020}. Polarizing beam splitters are used to decouple the $\SI{780}{\nano\meter}$ and $\SI{987}{\nano\meter}$ detection paths.
From the forward detection signal, we generate an error signal that is used to cool the particle to the millikelvin regime via feedback along all three motional axes \cite{dania2020}.
Both the feedback and minimization of excess micromotion localize the particle well below $\lambda/4$, enabling linear detection through self-homodyne.

\begin{figure}[ht]
	\centering
	\includegraphics[width=1\linewidth]{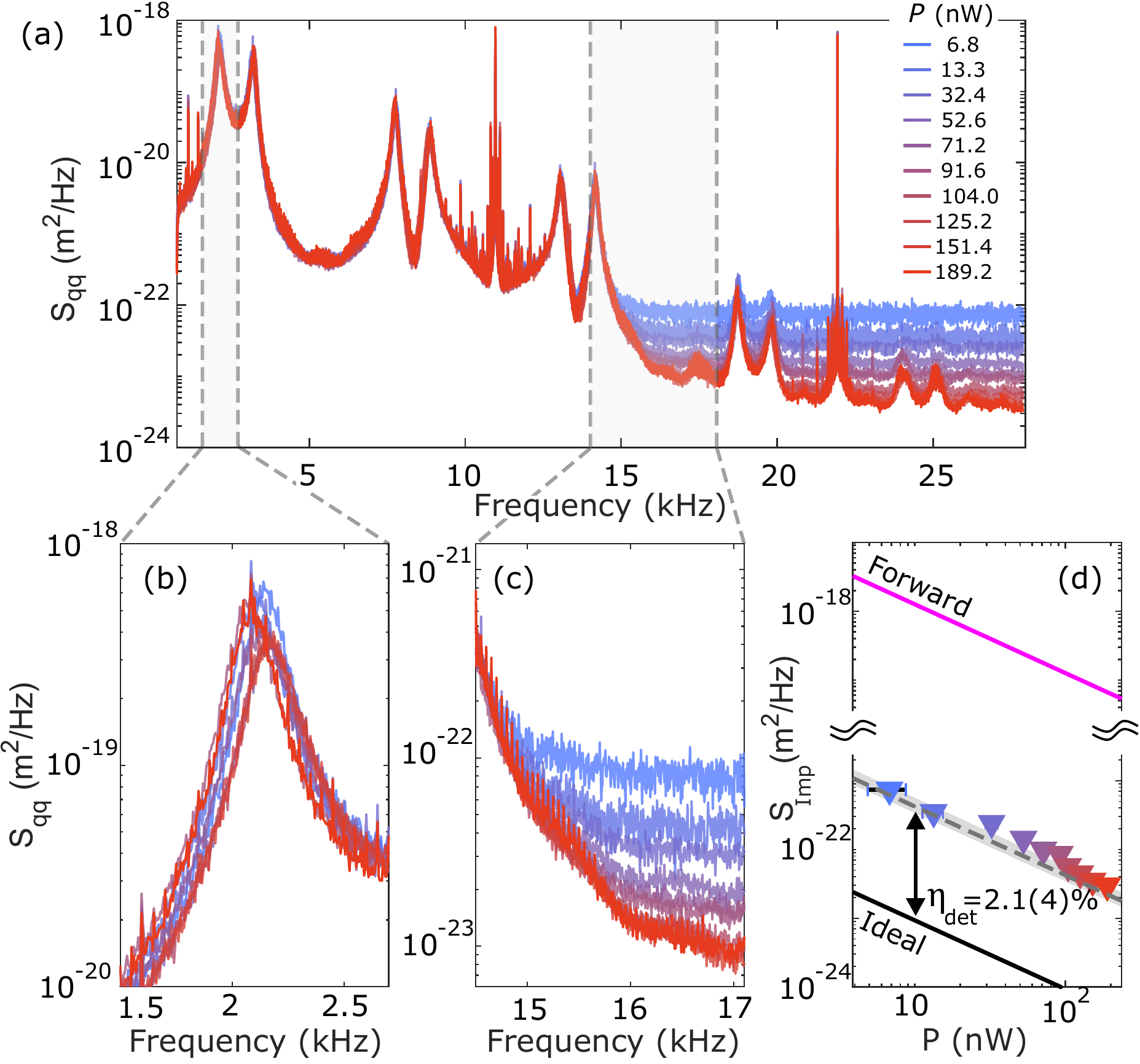}
	\caption{Motional spectra acquired with self-homodyne detection for different values of the total scattered power $P$ and with the same trapped particle.	(a) Spectra over a frequency range that includes all detected mechanical resonances. (b) Closer view of the first radial resonance.	(c) Closer view of the noise floor separation. Lower values of the noise floor are reached for higher values of the scattered power. Gain-dependent excess noise is subtracted.
	d) Imprecision $S_{\mathrm{imp}}$ of self-homodyne detection as a function of the scattered power $P$ of the particle, plotted as triangles. Error bars correspond to the uncertainty in the NA estimation. 
	The black solid line indicates the expected imprecision for ideal detection \cite{cerchiari2021} and the purple solid line indicates the imprecision for the forward detection inferred from our measurements. The gray dashed line and area represent the theoretically expected imprecision and its uncertainty, respectively. The distance between the black and gray lines is the detection efficiency $\eta_{\mathrm{det}}$. }

	\label{fig:fig_3}
\end{figure}

Implementing forward detection allows us not only to cool the particle but also to compare the detection efficiencies of the two methods.  For this comparison, two spectra are acquired simultaneously with the same particle. By fitting the calibrated self-homodyne spectrum with a Lorentzian, we extract the particle's temperature, which we use as a calibration for the spectrum obtained with the forward-detection method.

In Fig.~\ref{fig:fig_2}b, we plot the motional spectra acquired with the self-homodyne and forward detection techniques.  The two spectra have overlapping peaks at several frequencies that correspond to resonant motion of the particle.  However, the noise floor of the self-homodyne spectrum is $\SI{38}{\deci\bel}$ below that of the forward detection spectrum, and the new method resolves motional peaks that were hidden below the forward-detection noise floor. 
The noise floor of a calibrated motional spectrum corresponds to the measurement imprecision \cite{clerk2010}; the imprecision of the self-homodyne measurement is \SI{3.0(1)e-24}{\meter^2/\hertz}. 
The square root of this value corresponds to the motional sensitivity.
From calculations based on \cite{cerchiari2021}, we expect the ratio of the measurement imprecision for the two cases to be $\SI{26}{\deci\bel}$ for identical detection conditions.  
Taking into account the different powers and losses in the two optical paths, we predict a ratio of $\SI{32}{\deci\bel}$. We attribute the additional $\SI{6}{\deci\bel}$ to the particle not being centered in the focus of the forward beam~\cite{dania2020}.

Recently, measurement-based feedback cooling of nanoparticles to the motional ground state has been achieved in high-NA optical traps  via detection of backward-scattered light~\cite{magrini2021,tebbenjohanns2021}. The backward-detection technique is more efficient than forward detection, but nevertheless, from calculations based on Refs.~\cite{tebbenjohanns2019,cerchiari2021} at our working NA of 0.18, we expect the self-homodyne imprecision to be 0.58 times smaller than that achieved with backward detection. We refer to Ref.~\cite{cerchiari2021} for a detailed comparison of the three methods.

We now turn our attention to how our implementation of self-homodyne detection compares to an ideal detection scenario.  First, we characterize how the calibrated self-homodyne spectra depend on the scattered power over the full solid angle, $P$, which is tuned by adjusting the self-homodyne beam power. 
As this power is varied, we correct for particle displacements induced by radiation pressure forces by applying electric fields in order to preserve optical alignment.  For each value of $P$, we calibrate the motional spectrum by scanning the self-homodyne mirror.  The calibrated spectra are shown in Fig.~\ref{fig:fig_3}a; we see in Fig.~\ref{fig:fig_3}b that the spectra overlap at the particle's mechanical frequency and in Fig.~\ref{fig:fig_3}c that each spectrum approaches a unique noise floor at high frequencies.
	
The data of Fig.~\ref{fig:fig_3} show that increasing the power in the measurement field reduces the detection imprecision, as expected.  Increasing the power does not, however, shift the mechanical resonance shown in Fig.~\ref{fig:fig_3}b, which tells us that the role of the gradient force is negligible, consistent with the fact that the self-homodyne beam is nearly collimated. We also see in Fig.~\ref{fig:fig_3}b that the resonant peak height remains constant, from which we deduce that radiation-pressure shot noise is not heating the particle \cite{teufel2016} and thus that back action plays a negligible role at these laser powers and this pressure, as expected from calculations based on our system parameters (see Supplemental Material).
	
With these results in hand, we are now able to make a comparison with ideal detection. The detection imprecision is calculated as~\cite{cerchiari2021}
\begin{equation}
S_{\mathrm{imp}}=\frac{5\hbar c \lambda}{8\pi \eta_{\mathrm{det}} P},
\label{eq:1}
\end{equation}
where $\hbar$ is the reduced Plank constant, $c$ is the speed of light in vacuum, and $\eta_{\mathrm{det}}$ is the detection efficiency, which is a function of the NA, the interference visibility and the optical losses (see Supplemental Material). The working NA is fixed by the lenses used in the setup, which in our implementation are outside of the vacuum chamber; all the other relevant loss channels are independently measured. 
From these known losses, we calculate $\eta_{det}$ to be $0.021(4)$.  Including this value in Eq.~\ref{eq:1}, we predict the imprecision as a function of $P$, which agrees with our measured data, as shown in Fig.~\ref{fig:fig_3}d.  For comparison, in the same figure, we plot both the ideal case, corresponding to $\eta_{\mathrm{det}}=1$, and the imprecision of the forward technique extracted from our measurements.

\begin{figure}[ht]
	\centering
	\includegraphics[width=1\linewidth]{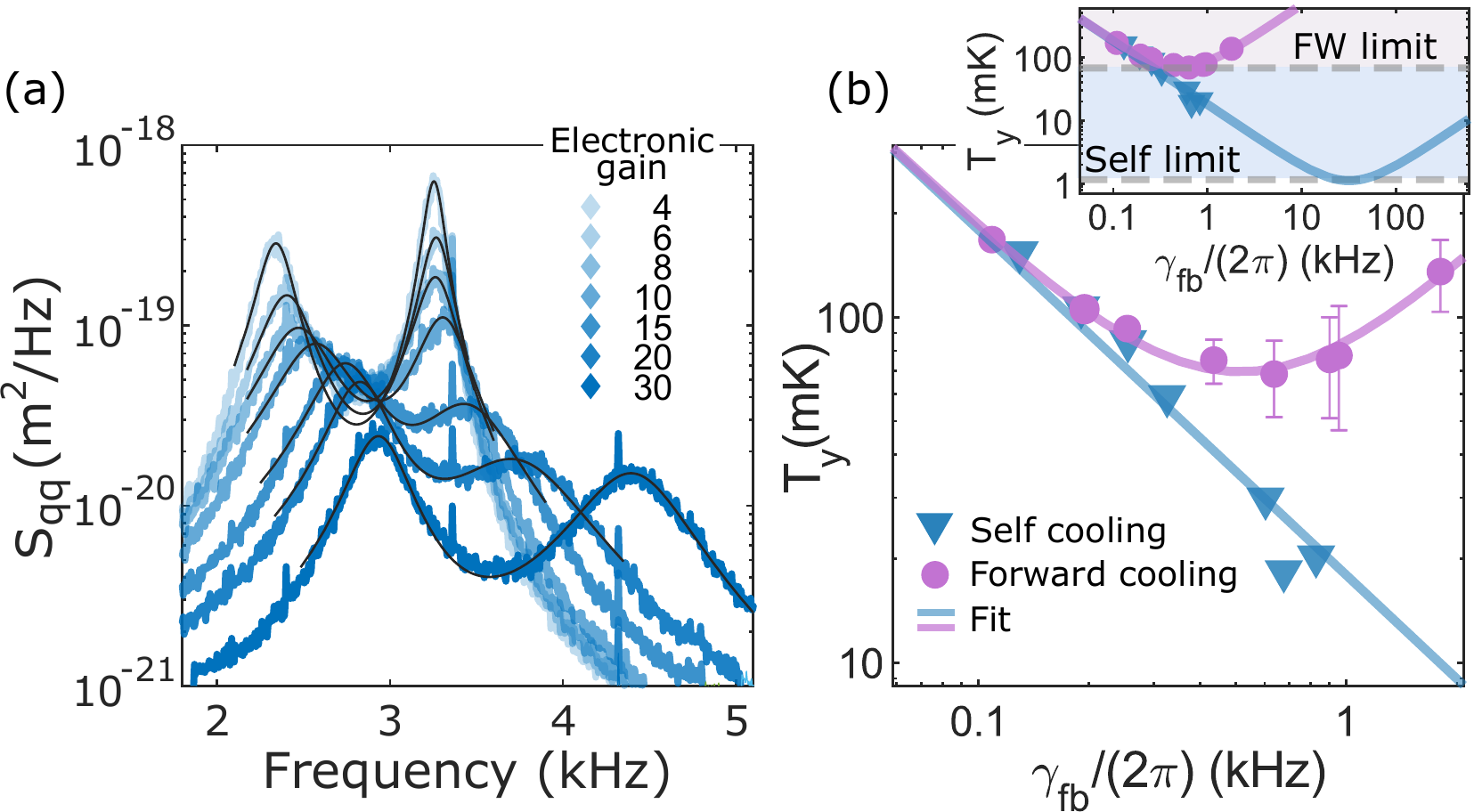}
	\caption{Feedback cooling of the particle's radial motion. (a) Motional spectra acquired with self-homodyne detection for different electronic feedback gains. Black lines are fits to Lorentzians, used to determine the temperature and cooling rate for each of the two modes. (b) Temperature $T_y$ of the higher-frequency radial mode as a function of the cooling rate $\gamma_{\mathrm{fb}}$ obtained with the self-homodyne (blue diamonds) and forward-detection methods (purple circles). In the self-homodyne case, temperatures are fitted with $T_y = A/\gamma_{\mathrm{fb}}$, where $A = \SI{112(7)}{\radian\kelvin}$ (blue line). In the forward-detection case, the temperature fit function includes the effect of measurement imprecision in the feedback loop \cite{dania2020}. In the inset, the latter fit function is applied to both temperature data sets.}
	\label{fig:fig_5}
\end{figure}

The improved detection efficiency with respect to forward detection motivates us to harness the self-homodyne method for feedback cooling of the particle.
We emphasize that this experiment was not designed to reach the ground state but rather as a proof-of-principle demonstration.
The output of the self-homodyne detector is filtered and phase shifted in order to generate a feedback signal, which is applied to a set of electrodes in order to displace the particle.  
To ensure that the particle is cooled primarily via the self-homodyne signal, we reduce the gain of the feedback cooling based on forward detection by $\SI{37.8}{\deci\bel}$ in the radial plane; the forward feedback alone cools the particle to $\SI{200}{\kelvin}$ in this plane.
Motion along $\hat{z}$ is not detected via self-homodyne, so we do not suppress cooling via forward detection along that axis.
In Fig.~\ref{fig:fig_5}a, we plot self-homodyne spectra for different values of the feedback strength.
As the feedback increases, the peak heights decrease due to the cooling force, and they shift to the right due to the spring force induced by the feedback (see Supplemental Material).

The temperatures $T_y$ achieved along the $\hat{y}$ axis via feedback cooling are deduced by fitting the calibrated spectra with a Lorentzian and are plotted in Fig.~\ref{fig:fig_5}b as a function of the cooling rate $\gamma_{\mathrm{fb}}$.
For comparison, in the same figure, we plot $T_y$ achieved with forward-detection feedback. We observe two distinct behaviors of the temperatures achieved with the two techniques: In the case of self-homodyne-based feedback, $T_y$ is proportional to $\gamma_{\mathrm{fb}}^{-1}$, yielding a minimum measured temperature of $\SI{18(1)}{\milli\kelvin}$. In the case of forward-detection feedback, $T_y$ exhibits a minimum that arises from the balance between cooling strength and measurement noise introduced by the feedback loop.  This minimum temperature is $3.8$ times the one reached with self-homodyne.

From a fit to the self-homodyne data and from the imprecision values determined from the calibrated spectra, we infer that for a higher value of the cooling rate, $\gamma_{\mathrm{fb}} = 2\pi\times\SI{31}{\kilo\hertz}$, self-homodyne-based cooling would reach a minimum temperature $T_{\mathrm{min}} = \SI{1}{\milli\kelvin}$~\cite{tebbenjohanns2019b}, corresponding to a phonon occupation number $\bar{n} = \SI{5e3}{}$, as shown in the inset of Fig.~\ref{fig:fig_5}b. This value is almost two orders of magnitude lower than the limit temperature achieved with forward cooling. We do not reach $T_{\mathrm{min}}$ here; \ this value of $\gamma_{\mathrm{fb}}$ lies in the overdamped regime ($\gamma_{\mathrm{fb}} \gg \nu_y$), where the filter electronics are not able to separate the radial modes effectively. Thus, we infer that noise limits the lowest achievable temperature.
We calculate that this noise has some contribution from pressure but is dominated by other sources, which may include fluctuations in the voltage on the trap electrodes, laser intensity fluctuations, or residual micromotion induced by the high $q$ parameter of the trap ($q=0.8$). 
Further measurements would be needed to distinguish between these sources.

In conclusion, we have realized self-homodyne detection of a nanoparticle's motion in a Paul trap and have demonstrated feedback cooling of that motion. The interferometric visibility between the nanoparticle and its image was measured to be $70\%$. By including a hemispherical mirror inside the vacuum chamber~\cite{higginbottom2018, araneda2020}, we expect to obtain a visibility close to unity, and by increasing the NA to $0.35$, we expect to reach the motional ground state ($\bar{n}<1$), provided that measurement back-action is the main source of decoherence. We thus anticipate that the self-homodyne method will overcome the obstacle of limited optical access in Paul traps\cite{dania2020,penny2021} as well as in hybrid traps\cite{millen2015,Conangla2020} and magnetic traps\cite{gieseler2020,tim2019,bradley}.  Given the low intrinsic decoherence in these traps due to the lack of photon scattering, we expect that this detection approach will enable quantum experiments with macroscopic objects with a unique degree of isolation. 
As a final remark, the self-homodyne scheme can be used to measure $q^2$ if the mirror's position is locked to the minimum of the interference fringe in Fig.~\ref{fig:fig_1}b~\cite{cerchiari2021b}. It has been pointed out that in an environment with sufficiently low decoherence, such a measurement would project a macroscopic wavefunction into a superposition state~\cite{oriol2011}.

\begin{acknowledgments}
We thank Felix Tebbenjohanns and Martin Frimmer for discussions about the calibration technique and Simon Baier and Oriol Romero-Isart for 
 their feedback on the manuscript.
This work was supported by Austrian Science Fund (FWF) Project No. Y951, by the ESQ Discovery grant “Sympathetic detection and cooling of nanoparticles levitated in a Paul trap” of the Austrian Academy of Sciences, by the European Union’s Horizon 2020 research and innovation program under the Marie Sk\l odowska-Curie grant agreement No 801110, and by the Austrian Federal Ministry of Education, Science and Research (BMBWF). It reflects only the authors' view; the EU agency is not responsible for any use that may be made of the information it contains. G. A. thanks Wolfson College, Oxford for support. 
\end{acknowledgments}

\appendix
\section{Position calibration}\label{appendix:calibration}
In the main text, we describe the calibration of the nanoparticle's amplitude of motion under the paraxial approximation, which is valid for a low numerical aperture (NA).  
This approximation allows us to assume that we will measure the same oscillation independent of whether it is the mirror or the scatterer that is displaced.
Here we quantify the extent to which our experimental parameters deviate from the paraxial approximation.
\begin{figure*}[ht]
	\centering
	\includegraphics[width=1\linewidth]{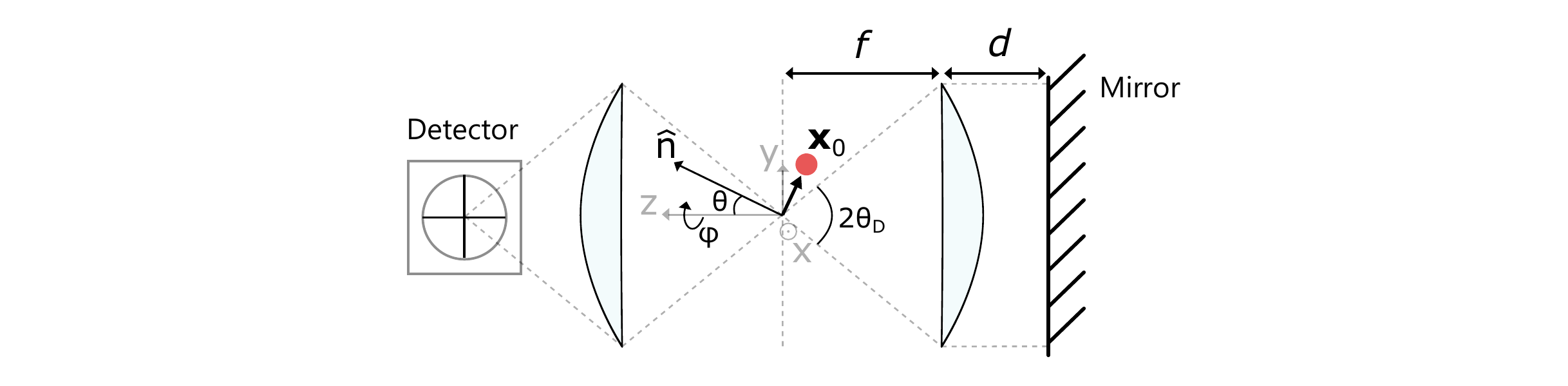}
	\caption{Coordinate system used in the derivation of the detector sensitivity. Here, $\bm{x}_0$ is the position of the dipolar scatterer, $\hat{n}$ is a generic direction in space, $\theta$ and $\phi$ are the angles of a spherical coordinate system defined by the map $(x, y, z) = r(\cos{\phi}\sin{\theta}, \sin{\phi} \sin{\theta}, \cos{\theta})$, $2\theta_D$ is the angle subtended by the lenses, $f$ is the focal length of both lenses, and $d$ is the distance between the flat mirror and the nearest lens.}
	\label{fig:supp_axes}
\end{figure*}

We use the coordinate system shown in Fig.~\ref{fig:supp_axes} to describe the setup. Note that, in contrast to the convention used in the main text, here the Cartesian coordinates $\{x,y,z\}$ no longer coincide with the axes of motion in the Paul trap, which we will not consider in this appendix. Our detector collects light over the solid angle $D$ defined by the angular intervals $\phi=[0, 2\pi)$ and $\theta=[0,\theta_D]$, with $\theta_D=\arcsin\left(\textrm{NA}\right)$. The normalized intensity $I_{\mathrm{tot}}$ at the detector depends on the position $\bm{x}_0$ of the scatterer \cite{cerchiari2021}:
\begin{align}
&I_{\mathrm{tot}} = \int_{D}\left(1+\rho^2-2\rho\cos{\psi }\right) dp,\label{eq:variation_of_I}\\
&\textrm{with} \quad \psi = \frac{4\pi}{\lambda}\left(\hat{\bm{n}} \cdot \bm{x}_0+R_s\right) \label{eq:variation_of_I_psi}\\
&\textrm{and} \quad dp = \frac{3}{8\pi}\left(1-\lvert\hat{\bm{\epsilon}}\cdot \hat{\bm{n}}\lvert^2\right) d\Omega.\label{eq:variation_of_I_dp}
\end{align}
Here, $\rho$ is the reflectivity of the electric field at the mirror, $dp$ is the differential dipole-radiated power, $\lambda$ is the wavelength, $\hat{\bm{n}}$ is a generic direction in space, $R_s = f + d$ is the sum of the lens focal length $f$ and the distance $d$ to the mirror, and  $\hat{\bm{\epsilon}}=\hat{\bm{y}}$ is the polarization direction of the illuminating field. 
Note that $R_s$ is the length of all optical paths from the focal point to the flat mirror because the lens collimates the outgoing spherical wavefronts into plane waves, such that any point of the mirror is equidistant from the focal point. Also note that the sum $1+\rho^2$ is constant and can be ignored when we evaluate the variation of the detected intensity induced by the nanoparticle's or the mirror's movements. Therefore, we limit our analysis to the term
\begin{equation}\label{eq:Ired}
I = -2\rho\int_{D}\cos{\psi}~dp.
\end{equation}
Inserting Eqs.~\ref{eq:variation_of_I_psi} and \ref{eq:variation_of_I_dp} into Eq.~\ref{eq:Ired} and using the angle addition formula, we obtain
\begin{widetext}
\begin{equation}\label{eq:is_split}
I = -2\rho \left(\cos{\left(\frac{4 \pi R_s}{\lambda}\right)} \int_D \cos{\left(\frac{4 \pi}{\lambda}\hat{\bm{n}} \cdot \bm{x}_0\right)} dp - \sin{\left(\frac{4 \pi R_s}{\lambda}\right)} \int_D \sin{\left(\frac{4 \pi}{\lambda}\hat{\bm{n}} \cdot \bm{x}_0\right)} dp\right),
\end{equation}
\end{widetext}
where the terms $\cos\left(\frac{4 \pi R_s}{\lambda}\right)$ and $\sin\left(\frac{4 \pi R_s}{\lambda}\right)$ have been brought in front of the integral sign because they have no dependence on $\hat{\bm{n}}$. In our experiments, the amplitude of the particle's motion is smaller than the wavelength. To first order in $|\textbf{x}_0|/\lambda$, the first integral in Eq.~\ref{eq:is_split} has a constant value independent of the particle and mirror positions, while the second term is zero for particle displacements in the $xy$--plane. We thus study how $I$ varies as a function of the particle's displacements along only the $z$ axis: We set $\bm{x}_0=(0,0,q)$ and substitute $\hat{\bm{n}} \cdot \bm{x}_0=q\cos\theta$ into Eq.~\ref{eq:is_split}.
Denoting the first integral on the right-hand side of Eq.~\ref{eq:is_split} as $a=\int_D\cos(\frac{4\pi}{\lambda}q\cos\theta) dp$ and the second integral as $b=\int_D\sin(\frac{4\pi}{\lambda}q\cos\theta) dp$, we use a trigonometric identity to rewrite $I$ in a more convenient form as 
\begin{align}
&I=-A\cos{\left(\frac{4\pi R_s}{\lambda}+\phi\right)}\text{, where}\label{eq:I_sin}\\
&A=2\rho\sqrt{a^2+b^2}\quad\text{and}\quad\phi=\arctan\left(\frac{b}{a}\right).\label{eq:phi_A}
\end{align}
Equation \ref{eq:I_sin} shows us that the intensity $I$ at the detector varies sinusoidally as a function of the optical path $R_s$, which in turn depends linearly on the mirror--lens distance $d$. Note that $a$ and $b$, and therefore the amplitude $A$ and the phase $\phi$, are functions of $q$ and $\theta_D$ that do not depend on the optical path $R_s$.
\subsection{Mirror displacement}
Expanding the expression for the interference amplitude $A$ from Eq.~\ref{eq:phi_A} up to second order in the collection angle $\theta_D$ yields
\begin{equation}\label{eq:A_mirror}
A=\frac{3\rho\theta_D^2}{4},
\end{equation}
the value of which does not depend on the particle position $q$. We remind the reader that the lens--mirror distance $d$ enters in the optical path as $R_s=f+d$, where $f$ is the fixed focal length of the lens. The detector sensitivity to mirror displacements is thus expressed as
\begin{equation}\label{eq:chi_m}
\chi_m=\text{max}\left(\frac{\partial I}{\partial R_s}\right)=\frac{4\pi A}{\lambda},
\end{equation} 
which does not depend on the particle position.
\subsection{Particle displacement}
Information about the particle position is encoded in the interference phase $\phi$ of Eq.~\ref{eq:I_sin}. Expanding the expression for $\phi$ from Eq.~\ref{eq:phi_A} up to second order in $\theta_D$ yields
\begin{equation}\label{eq:phi}
\phi=\frac{4\pi q}{\lambda}\left(1-\frac{\theta_D^2}{4}\right),
\end{equation} 
from which we see that the phase $\phi$ is linearly proportional to the particle position $q$. To read out $q$, we stabilize the lens--mirror distance $d$ so as to satisfy the relation $R_s=f+d=\frac{\lambda}{8}(2n+1)$, where $n$ is an integer. This condition ensures that for $q\ll\lambda$, the particle motion is mapped linearly onto the intensity $I$ with maximum sensitivity.
Substituting this value for $R_s$ into Eq.~\ref{eq:I_sin} together with the expression for $\phi$ found in Eq.~\ref{eq:phi}, and neglecting terms $O(\theta_D^4)$, we obtain an expression for $I$ as a sinusoidal function of $q$, 
\begin{equation}
I=-A\sin{\left(\frac{4\pi q}{\lambda}\left(1-\frac{\theta_D^2}{4}\right)\right)},
\end{equation}
from which we determine the maximum detector sensitivity to particle displacements:
\begin{equation}\label{eq:chi_p}
\chi_p=\text{max}\left(\frac{\partial I}{\partial q}\right)=\frac{4\pi A}{\lambda}\left(1-\frac{\theta_D^2}{4}\right).
\end{equation}
Here again, we have used the fact that, to second order in $\theta_D$, the amplitude $A$ is independent of $q$. 

\begin{figure*}
	\centering
	\includegraphics[width=0.5\linewidth]{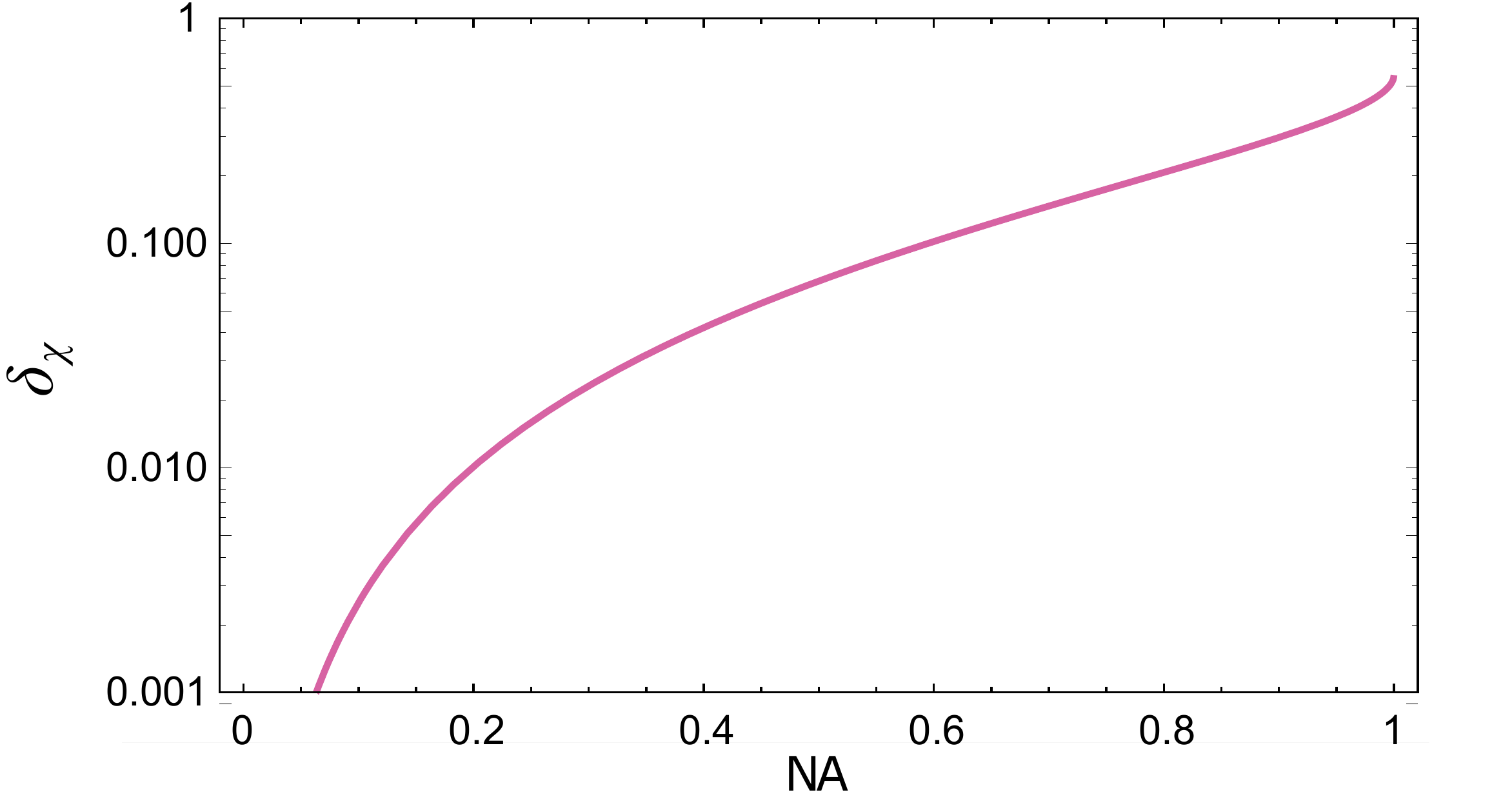}
	\caption{Relative deviation $\delta_\chi$ of the calibration measurement from the sensitivity to position displacement of the dipolar scatterer as a function of the system numerical aperture (NA).}
	\label{fig:calibration}
\end{figure*}

In the main text, the calibration of the self-homodyne detection method assumes that $\chi_p=\chi_m$. Equations~\ref{eq:chi_p} and \ref{eq:chi_m} now allow us to quantify the error of this approximation. In our setup, $\text{NA}=\sin(\theta_D)=0.18$, for which we extract the relative difference between the mirror and particle sensitivities
\begin{equation}
\delta_{\chi}=2\frac{\chi_m-\chi_p}{\chi_m+\chi_p}=0.008\,,
\end{equation}
which validates our approximation. Figure~\ref{fig:calibration} shows the value of $\delta_{\chi}$ calculated using the full analytical solution of Eq.~\ref{eq:is_split} for values of the numerical aperture $0<\text{NA}<1$.  We see that the deviation remains below $10\%$ if the condition $\text{NA}\lesssim0.6$ is satisfied.

\section{Power calibration}
\begin{figure*}[ht]
	\centering
	\includegraphics[width=1\linewidth]{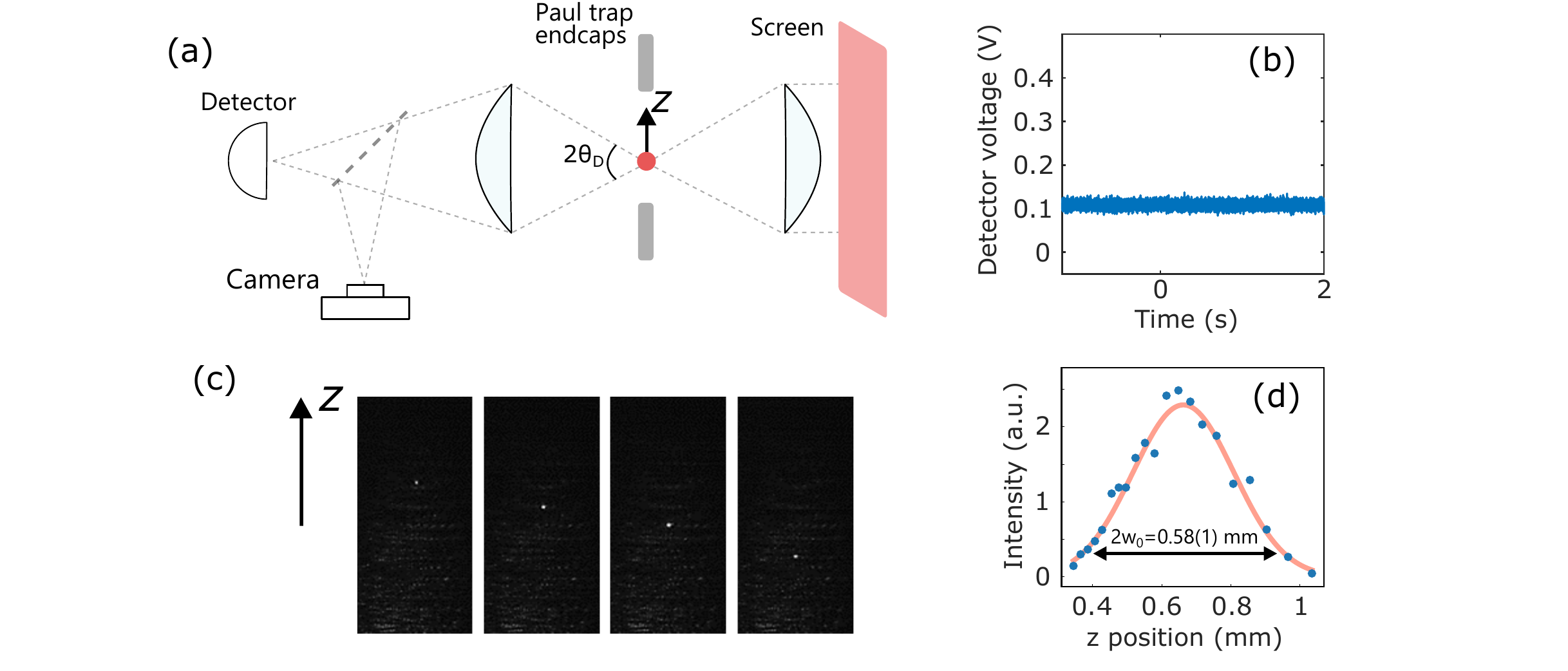}
	\caption{Procedure for determining the total scattered power $P$ of the particle. (a) Sketch of the setup. A laser beam linearly polarized along the $z$ axis illuminates the nanoparticle, and a lens with numerical aperture $\text{NA}=\arcsin(\theta_D)=0.18$ collects the scattered light. The mirror is blocked, so only one image is formed at the detector. (b) Detector time trace obtained with the mirror covered by the screen. (c) Camera images for different positions of the particle along the Paul trap axis $z$. (d) Intensity of the nanoparticle's image on the camera as a function of the particle's position along the $z$ axis. The orange line is a Gaussian fit, from which we extract the waist $w_0$ and its uncertainty.}
	\label{fig:supp_waist}
\end{figure*}

In this section, we calculate the nanoparticle's scattered light power $P$ that has been used in Figs.~3a and 3d in the main text to extract the self-homodyne detection efficiency. The setup for a power measurement is illustrated in Fig.~\ref{fig:supp_waist}a. The particle is illuminated with a laser beam linearly polarized along the $z$ axis, and a fraction of the scattered power is collected by a lens with numerical aperture $\text{NA}=0.18$. The total scattered power $P$ is obtained from
\begin{equation}\label{eq:P_scatt}
P=\frac{P_{\mathrm{col}}}{\eta_{\mathrm{col}}},
\end{equation}
where $\eta_{\mathrm{col}}$ and $P_{\mathrm{col}}$ are the lens collection efficiency and collected power, respectively. We calculate $\eta_{\mathrm{col}}$ as
\begin{equation}
\eta_{\mathrm{col}}=\frac{\int_{0}^{\theta_D}dP}{\int_{\Omega}dP}=0.012,
\end{equation}
where $\theta_D=\arcsin(\text{NA})$ and $dP$ is the dipole radiation emission per unit solid angle $\Omega$. Note that we have modeled the nanoparticle as a dipolar scatterer since the particle's radius $r=\SI{150}{\nano\meter}$ is smaller than the wavelength $\lambda=\SI{780}{\nano\meter}$.
The power $P_{\mathrm{col}}$ is directly measured by a calibrated detector. During a measurement of $P_{\mathrm{col}}$, the mirror is blocked in order to avoid interference with the nanoparticle's image that would lead us to overestimate or underestimate the collected power. A typical detector output is shown in Fig.~\ref{fig:supp_waist}b: no interference is observed, and the noise is dominated by shot noise. For an illumination beam power of $\SI{430}{\milli\watt}$, we find $P=\SI{84(10)}{\nano\watt}$, where the error is due to the uncertainty in the NA estimation. Eq.~\ref{eq:P_scatt} has been used to extract the values of $P$ in the main text.

As an independent estimation of $P$, we calculate the total power scattered by a dielectric sphere with radius $r$ and index of refraction $n_p$ illuminated by a laser beam with intensity $I_0$, using the Rayleigh approximation. This power is
\begin{equation}
P=I_0\sigma,
\label{eq:P_rey}
\end{equation} 
where $\sigma=\frac{8\pi}{3}(\frac{\alpha_p k^2}{4\pi\epsilon_0})^2$ is the scattering cross section, with $k=2\pi/\lambda$ the wave number, $\alpha_p=4\pi r^3\epsilon_0\frac{(n^2_p-1)}{(n^2_p+2)}$ the particle's polarizability, and $\epsilon_0$ the vacuum permittivity.  Here, $I_0=\frac{2P_0}{\pi w_0^2}$, with $P_0$ and $w_0$ the beam power and electric-field waist ($1/e$ radius), respectively.
We determine the beam waist from a measurement of the intensity of the particle's image as its position with respect to the beam is varied along the $z$ axis. This procedure is illustrated in Fig.~\ref{fig:supp_waist}(c), in which we see a sequence of images of the particle at different positions, obtained with a CMOS camera. The images are calibrated using the known distances between the electrodes.  We extract $2w_0=\SI{0.58(1)}{\milli\meter}$ from a Gaussian fit to the intensity profile (Fig.~\ref{fig:supp_waist}(d)). According to the manufacturer, the particle size and index of refraction are $r=\SI{150(15)}{\nano\meter}$, and $n_p=1.45$, respectively. By inserting these values into Eq.~\ref{eq:P_rey} along with a beam power $P_0=\SI{430}{\milli\watt}$, we find a total scattered power $P=\SI{0.09(5)}{\micro\watt}$, in agreement with the APD measurement. The large error in estimating $P$ with Eq.~\ref{eq:P_rey} follows from the large uncertainty in the particle's radius and from the fact that $P\propto r^6$.

\section{Detection imprecision}
\label{app_c}
The self-homodyne detection imprecision for displacements along the particle--mirror axis $q$ is given in Eq.~1 of the main text.  The detection efficiency $\eta_{\mathrm{det}}$ in that equation is a function of the NA, the interferometric visibility $\mathcal{V}$, and optical losses. It can be written as~\cite{cerchiari2021}
\begin{widetext}
\begin{equation}
\eta_{\mathrm{det}}=\eta_{\mathrm{\mathcal{V}}}\eta_{\mathrm{opt}}\eta_{\mathrm{QE}}\frac{(128-90\cos{\theta_D}-35\cos{3\theta_D}-3\cos{5\theta_D})}{128},
\end{equation}
\end{widetext}
where we have decomposed the efficiency into  a product of the loss sources. Here $\theta_D=\arcsin{0.18}$ is our lens half-aperture angle, $\eta_{\mathrm{\mathcal{V}}}=\mathrm{\mathcal{V}}^2 = 0.7^2$ expresses the losses due to mode mismatch,  $\eta_{\mathrm{opt}}=0.9$ represents the optical path losses, and $\eta_{\mathrm{QE}}=0.82$ is the quantum efficiency of our detector. Note that, as shown in Fig.~\ref{fig:supp_fb}a, the angle between the detection axis~$q$ and each of the two radial trap axes is $45^{\circ}$, so the imprecision in detecting the motion along these axis accrues an additional factor of 2.  This factor is included in Eq.~1 of the main text.

\section{Micromotion minimization}
We minimize excess micromotion with three techniques that have been adapted 
from methods used in experiments trapping atomic ions~\cite{berkeland1998,keller2015}.  In all three cases, we control the nanoparticle's position in the trap's radial plane by applying DC voltages to compensation electrodes.

The first, rough method is to minimize the height of the micromotion peak in the motional spectrum --- that is, the peak corresponding to motion at the trap drive frequency --- while the particle's position in the trap is varied. 
The second, more accurate compensation step is to observe correlations between the phase of the AC trap voltage and the phase of the particle's motion at the trap drive frequency. When the particle is moved across a null line of the trap field, the phase of the particle's motion at the trap drive frequency jumps by $\pi$ with respect to the phase of the AC field applied to the electrodes. Thus, excess micromotion along one trap axis is minimized when the particle is placed at the phase-jump point. The procedure is repeated along two orthogonal axes in the trap's radial plane. We measure the particle's micromotion by filtering the signal from the forward detector at the trap drive frequency, and at the same time, we monitor the trap voltage with a pick-off signal from the trap's electrodes.

In the last and most precise method, we scan the mirror position and observe interference fringes at the detector. If micromotion is not minimized, the fringe contrast is reduced. Therefore, we translate the particle in the radial plane and find a position that maximizes the fringe contrast. A typical procedure for minimizing micromotion is to apply the methods in the order described above.

\section{Feedback-induced axis rotation}	
Feedback forces that are applied to cool the nanoparticle
result in a rotation of the trapping potential with respect to the detection axis $q$. A wrong assumption about the motional axes' orientation could lead to overestimation or underestimation of the detected amplitude of motion, and thus, of the particle's temperature under feedback. Here we show how we correct the power spectral densities (PSDs) of the particle's motion to account for this effect.

In the absence of feedback, the radial motion of a nanoparticle in the Paul trap is described as two uncoupled harmonic oscillators, which obey the Hamiltonian
\begin{equation} \label{eq:hamiltonian}
H(x,y) = \frac{1}{2} m (\dot x ^2 + \dot y^2) + \frac{1}{2} m (\omega_x^2 x^2 + \omega_y^2 y^2), 
\end{equation}
where $\omega_x$ and $\omega_y$ are the radial mode frequencies and $m$ denotes the mass of the particle. In our experiments, we apply a DC bias to a pair of AC trap electrodes to break the trap symmetry and lift the degeneracy of the radial frequencies, so that $\omega_{x}\neq\omega_{y}$.

\begin{figure*}[ht]
	\centering
	\includegraphics[width=0.6\linewidth]{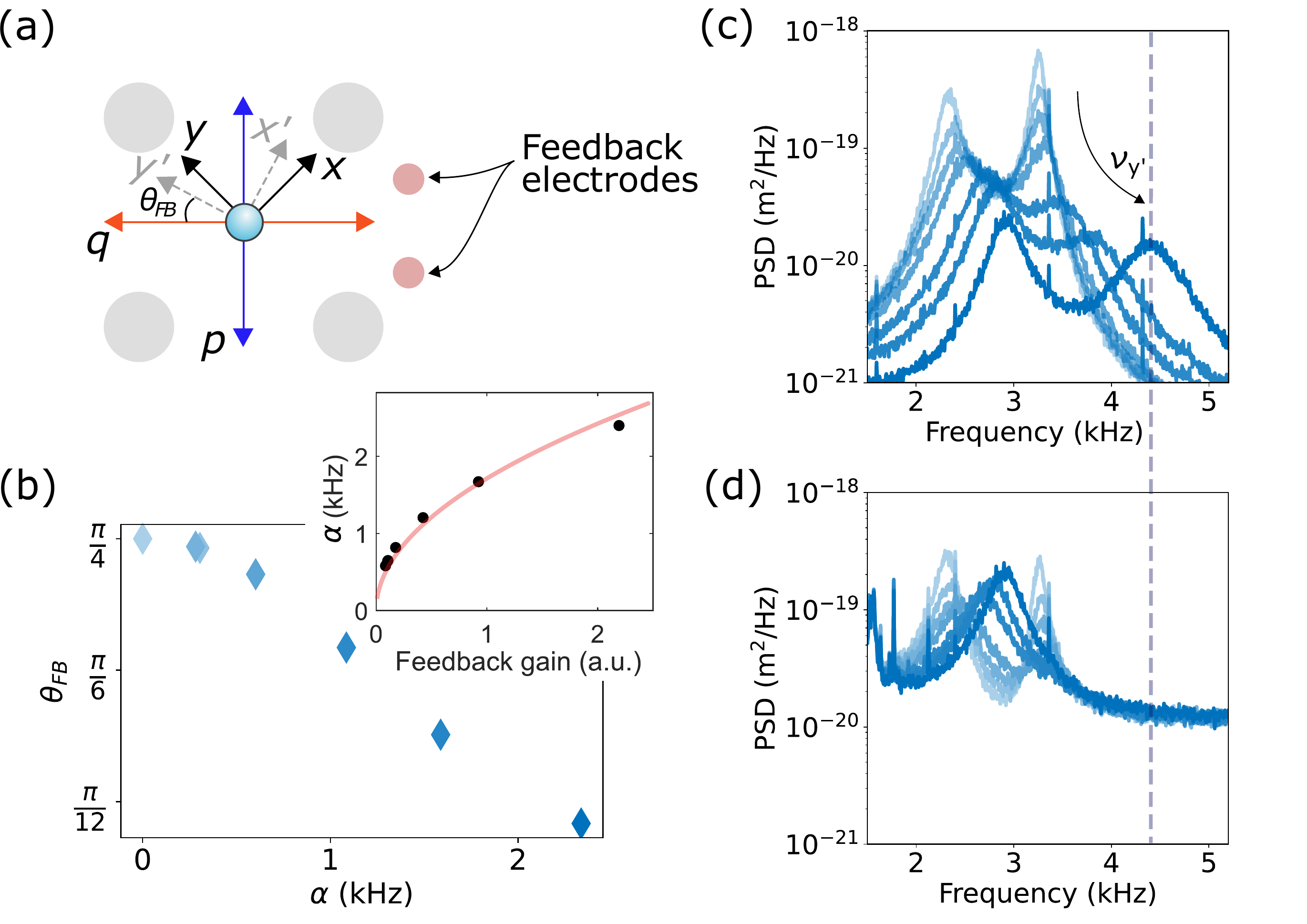}
	\caption{(a) The feedback force that cools the particle also leads to new eigenvectors of the particle's motion, which are no longer parallel to the $x$ and $y$ axes of the Paul trap.  The new eigenvectors are $\hat{x}'$ and $\hat{y}'$, and $\hat{y}'$ subtends an angle $\theta_{\mathrm{FB}}$ with respect to the self-homodyne detection axis $q$. The axis $p$ is the forward-detection axis. (b) The angle $\theta_{\mathrm{FB}}$ as a function of the gain parameter $\alpha$.  In the inset, $\alpha$ is plotted as a function of the electronic feedback gain and fitted with the equation $\alpha=A\sqrt{x}$, where $A=2700\pm200$. (c) The power spectral density (PSD) of the radial motion acquired with the self-homodyne method and (d) the forward-detection method for the same gain settings as in (b). The vertical dashed line indicates the highest eigenfrequency $\nu_{y'}$ reached along the $y'$ axis, as $\theta_{\mathrm{FB}}\to0$.  Motion at this frequency is not detected with the forward-detection method.}
	\label{fig:supp_fb}
\end{figure*}

To identify the orientation of the principal axes and the motional frequencies along these axes, it is useful to write the potential energy in matrix form:
\begin{equation}
U=\frac{1}{2} m
\begin{pmatrix}
x \\ y 
\end{pmatrix}^\intercal
\begin{pmatrix}
\omega_x^2  & 0\\
0 & \omega_y^2\\
\end{pmatrix}
\begin{pmatrix} x \\ y \end{pmatrix}.
\end{equation}
The matrix $U$ is diagonal, so we can directly read off the eigenvalues, which are proportional to $\omega_x^2$ and $\omega_y^2$, as well as the eigenvectors $\hat x =\begin{pmatrix} 1 \\ 0 \end{pmatrix} $ and $\hat y =\begin{pmatrix} 0 \\ 1 \end{pmatrix}$, which represent the Paul trap radial axes. Note that each radial axis connects opposing trap electrodes~\cite{Ghosh1995}, as shown in Fig.~\ref{fig:supp_fb}a.

Now we include feedback. In addition to a viscous cooling force, feedback produces a spring-like force~\cite{genes2008} that modifies the stiffness of the trap. Feedback is applied along the detection axis $q$ by means of two electrodes. 
As discussed in Appendix~\ref{app_c}, the angle between $\hat{q}$ and each of the two radial trap axes is $45^{\circ}$, as shown in Fig.~\ref{fig:supp_fb}(a).  As a result, the spring force shifts both mode frequencies, and the orientation of the eigenvectors rotates in the radial plane. 

We model the spring force by adding the term $\frac{1}{2} m \alpha^2 (x+y)^2 $ to the Hamiltonian \ref{eq:hamiltonian}, where the spring constant $m\alpha^2$ is proportional to the feedback gain. The matrix representation for the potential including the spring force now contains off-diagonal terms:
\begin{equation}
U = \frac{1}{2} m
\begin{pmatrix}
x \\ y 
\end{pmatrix}^\intercal
\begin{pmatrix}
\omega_x^2 + \alpha^2 & \alpha^2 \\ 
\alpha^2 & \omega_y^2 + \alpha^2 \\
\end{pmatrix}
\begin{pmatrix}
x \\ y 
\end{pmatrix}.
\label{eq:U}
\end{equation}
We diagonalize the $2\times2$ matrix in Eq.~\ref{eq:U} to find the new eigenfrequencies $\nu_{x'}, \nu_{y'}$ and eigenvectors $x',y'$:
\begin{widetext}
	\begin{align}
	\label{eq:alpha} 
	\nu_{x',y'}^2 &= \frac{1}{2}\bigg( 2\alpha^2 +\omega_x^2 +\omega_y^2 \mp \sqrt{4\alpha^4 +\omega_x^4 -2\omega_x^2\omega_y^2 + \omega_y^4}\bigg), \\
	x' &= \begin{pmatrix} c_{x} \\ 1 \end{pmatrix},~ y' = \begin{pmatrix} c_{y} \\ 1 \end{pmatrix}, \mathrm{where~} 
	c_{x,y} = \frac{1}{\alpha^2}(\nu_{x',y'}^2 - \omega_y^2)-1\,.
	\label{eq:cxy}
	\end{align}
\end{widetext}
Note that for $\alpha\to 0$, i.e., in the absence of feedback, the eigenfrequencies become the bare trap frequencies $\omega_{x}$ and $\omega_{y}$.  We identify the eigenfrequencies $\nu_{x'}$ and $\nu_{y'}$ from the measured PSDs. By inserting them into Eq.~\ref{eq:alpha}, we determine the value of $\alpha$ for each feedback gain setting, which we then use in Eq.~\ref{eq:cxy} to determine the eigenvectors.

We calculate the angle $\theta_{\mathrm{FB}}$ that  $y'$ subtends with respect to the self-homodyne detection axis $q$. The power spectral density of the radial motion $S_{qq}$ is then calculated as
\begin{equation}
S_{qq} = S_{x'x'} \sin^2{\theta_{\mathrm{FB}}}  + S_{y'y'}\cos^2{\theta_{\mathrm{FB}}}\,,
\label{eq:Sqq}
\end{equation}
where $S_{x'x'}$ and $S_{y'y'}$ are the single-sided PSDs of the particle's oscillation along $x'$ and $y'$. Finally, the values for the particle's temperature along the $y'$ axis given in the main text are calculated using the expression
\begin{equation}
T_{y'}=\frac{m\nu_{y'}^2\langle q^2\rangle}{k_B}\cdot\frac{1}{\cos^2(\theta_{\mathrm{FB}})},
\end{equation}
where $m=\SI{2.0(2)e-17}{\kilogram}$ is the particle mass measured by inducing particle-charge jumps~\cite{dania2020}, $k_B$ is the Boltzmann constant, and $\langle q^2\rangle$ is the measured motional variance, which is extracted from a fit of the calibrated spectra with a Lorentzian function~\cite{Hebestreit2018}.

Figure~\ref{fig:supp_fb}(b) shows $\theta_{\mathrm{FB}}$ as function of $\alpha$. 
Figures~\ref{fig:supp_fb}(c)--(d) show PSDs of the particle motion under feedback cooling via self-homodyne detection, as measured using forward and self-homodyne detection over a range of feedback gain settings. Increasing the feedback gain rotates the $y'$ axis towards the $q$ axis. As a consequence, the forward detection method, which is sensitive to particle displacements along the $p$ axis of Fig.~\ref{fig:supp_fb}(a), becomes less sensitive to motion along the $y'$ axis. For the largest gain setting, the peak corresponding to the $y'$ mode vanishes under the noise floor of the forward-detection PSD. 

\section{Radiation-pressure back-action}
Radiation-pressure shot noise from scattered light generates a back-action force on the particle, which for the highest power of the illumination beam used in our experiments yields a single-sided PSD of~\cite{tebbenjohanns2019} 
\begin{equation}
S_{\mathrm{BA}}=\frac{4}{5}\frac{\hbar kP}{c}=\SI{4e-43}{\newton^2/\sqrt{\hertz}},
\end{equation}
where $c$ is the speed of light in vacuum, $\hbar$ is the reduced Planck constant, and $k=2\pi/\lambda$. Another source of noise for the particle's motion is random collisions with the background gas at room temperature ($T=\SI{300}{\kelvin}$). This process has a force PSD given by
\begin{equation}
S_{\mathrm{gas}}=4k_BT\gamma m=\SI{4.6e-41}{\newton^2/\sqrt{\hertz}},
\end{equation}
The linewidth $\gamma=2\pi\times\SI{8.6}{\micro\hertz}$ accounts for gas damping and is calculated for our background pressure of \SI{2e-8}{\milli \bar} from a measured value of $\gamma=2\pi\times\SI{4.3}{\hertz}$ at a pressure of $\SI{1e-2}{\milli\bar}$, using the fact that under our experimental conditions, $\gamma$ scales linearly with pressure~\cite{bykov2019}. With these values, $S_{\mathrm{gas}}$ is two orders of magnitude larger than $S_{\mathrm{BA}}$, meaning that we are still far from the back-action-dominated regime, given our system parameters. 

\section{Fits to feedback-cooling data}
In this section, we provide more detail on the fits in Fig. 4b of the main text. For large values of the feedback-cooling rate $\gamma_{\text{fb}}$ compared to the natural damping rate $\gamma_0$, the steady-state temperature $T_y$ as a function of the cooling rate $\gamma_{\text{fb}}$ is~\cite{dania2020,tebbenjohanns2019b}
\begin{gather}\label{eq:TvsG}
T_y=\frac{A}{\gamma_{\text{fb}}}+B\gamma_{\text{fb}}\,,\,\,\text{where}\\
A=\gamma_0T_0\,\,\,\text{and}\,\,\,B=\frac{\pi m\omega_y^2}{2k_B}S_{\text{imp}}\,. \label{eq:TvsG2}
\end{gather}
Here $T_0$ is the equilibrium effective temperature of the thermal bath and $S_{\text{imp}}$ is the position measurement imprecision. 
Equation~\ref{eq:TvsG} yields a minimum temperature 
\begin{align}
T_{\text{min}}&=2\sqrt{AB}\,,\,\, \text{at the cooling rate}\label{eq:Tmin}\\
\gamma_{\text{fb}}^{\text{min}}&=\sqrt{\frac{A}{B}}\,.
\end{align}
In the case of self-homodyne-based feedback, we have data for $T_y$ as function of $\gamma_{\text{fb}}$ only for $\gamma_{\text{fb}}\ll\gamma_{\text{fb}}^{\text{min}}$, so it is not possible to measure $T_{\text{min}}$ directly. In this regime, the first term on the right-hand side of Eq.~\ref{eq:TvsG} dominates over the second term. Accordingly, we fit our data with the equation 
\begin{equation}
T_y=\frac{A}{\gamma_{\text{fb}}},
\end{equation}
where $A$ is the fit parameter, as shown in Fig. 4b of the main text. To estimate the minimum temperature achievable with feedback cooling based on self-homodyne detection, we solve Eq.~\ref{eq:Tmin} with the value of $A=\SI{112(7)}{\radian\kelvin}$ from the fit and the value of $B$ calculated from Eq.~\ref{eq:TvsG2}.
The values $\omega_{y}=2\pi\times\SI{3.2}{\kilo\hertz}$ and $S_{\text{imp}}=\SI{3e-24}{\meter^2/\hertz}$ are extracted from a calibrated spectrum.
We obtain $T_{\text{min}}=\SI{1}{\milli\kelvin}$.

In the case of forward-detection-based feedback cooling, the data for $T_y$ as a function of $\gamma_{\text{fb}}$ reach a minimum, so we fit these data with Eq.~\ref{eq:TvsG}, with both $A$ and $B$ as fit parameters, as shown in Fig. 4b of the main text. We obtain $A=\SI{105(20)}{\radian\kelvin}$ and $B=\SI{6.5(7)e-5}{\kelvin/\radian^2}$.

\bibliography{self_homodyne}

\end{document}